\documentclass[fleqn,10pt]{wlscirep}
\usepackage[utf8]{inputenc}
\usepackage[T1]{fontenc}
\title{Observation of quasi-ballistic thermal transport of surface phonon-polaritons over hundreds of micrometres}

\author[1,*]{Yunhui Wu}
\author[1,2]{Jose Ordonez-Miranda}
\author[1,2]{Laurent Jalabert}
\author[1]{Saeko Tachikawa}
\author[1]{Roman Anufriev}
\author[1]{Hiroyuki Fujita}
\author[1,2]{Sebastian Volz}
\author[1,2,*]{Masahiro Nomura}
\affil[1]{Institute of Industrial Science, The University of Tokyo, Tokyo 153-8505, Japan}
\affil[2]{LIMMS, CNRS-IIS IRL 2820, The University of Tokyo, Tokyo 153-8505, Japan}
\affil[*]{Corresponding authors. Email: yunhui@iis.u-tokyo.ac.jp (Y.W.)
nomura@iis.u-tokyo.ac.jp (M.N.)}


\begin{abstract}
Long-distance propagation of heat carriers is essential for efficient heat dissipation in microelectronics. 
However, in dielectric nanomaterials, the primary heat carriers -- phonons -- can propagate ballistically only for hundreds of nanometres, which limits their heat conduction efficiency. 
Theory predicts that surface phonon-polaritons (SPhPs) can overcome this limitation and conduct heat without dissipation for hundreds of micrometres. 
In this work, we experimentally demonstrate such long-distance heat transport by SPhPs. 
Using the 3$\omega$ technique, we measure the in-plane thermal conductivity of SiN nanomembranes for different heater-sensor distances (100 and 200 $\mu$m), membrane thicknesses (30 -- 200 nm), and temperatures (300 -- 400 K). 
We find that in contrast with thick membranes, thin nanomembranes support heat conduction by SPhPs, as evidenced by an increase in the thermal conductivity with temperature.
Remarkably, the thermal conductivity measured 200 $\mu$m away from the heater are consistently higher than that measured 100 $\mu$m closer.
This result suggests that heat conduction by SPhPs is quasi-ballistic over at least hundreds of micrometres.
Thus, our findings show that SPhPs can enhance heat dissipation in polar nanomembranes and find applications in thermal management, near-field radiation, and polaritonics.
\end{abstract}

\begin{document}
\flushbottom
\maketitle
\thispagestyle{empty}

\section*{INTRODUCTION}

Heat conduction in semiconductors is essentially governed by the mean free path of heat carriers. 
In nanomaterials of the size comparable to the mean free path, heat conduction can even happen ballistically, without dissipation, which could be used for heat dissipation in microelectronics \cite{pop2010energy}. 
However, even in silicon membranes, the mean free path of phonons at room temperature is as short as 30 -- 300 nm \cite{Anufriev2020}, which only becomes shorter at higher temperatures. 
Such short propagation length makes it unfeasible to attain the ballistic regime in the micro-sized dielectric materials used in modern electronics.
Therefore, heat carriers with a longer propagation length are desirable.

Over the past two decades, the propagation, detection, and energy transport of surface electromagnetic waves has attracted attention due to the predominance of surface effects over the volumetric ones in nanostructures with high surface-to-volume ratios \cite{yang_long-range_1991}. 
Certain types of surface electromagnetic waves may even carry heat \cite{joulain_surface_2005,chen_measurement_2007,chen_heat_2010,baffou_mapping_2010} and thus improve the thermal performance and stability of nanoscale devices \cite{mulet_nanoscale_2001,dusastre_materials_2017,biswas_high-performance_2012}. 
One type of such waves is the surface phonon-polaritons (SPhPs) -- evanescent surface waves generated by the coupling of photons with optical phonons \cite{chen_surface_2005, ordonez-miranda_anomalous_2013, greffet_coherent_2002,francoeur_local_2010}. 
Theoretical works predict that SPhPs can propagate along the surface of polar dielectric materials without dissipation for hundreds of micrometres \cite{ordonez-miranda_thermal_2014, nano10071383}, which is orders of magnitude longer than the typical mean free path of phonons. 
Moreover, recent experiments demonstrated that in thin membranes, SPhPs can even carry heat at least as efficiently as phonons \cite{wu2020enhanced,tranchant_two-dimensional_2019}.
However, the long-distance ballistic conduction remains to be experimentally demonstrated.

In this work, we aim to demonstrate ballistic heat transport via SPhPs over hundreds of micrometres.
Our experiments compare the thermal conductivity of SiN membranes measured 100 and 200 $\mu$m away from the heater.
Assuming that phonon transport in amorphous membranes is purely diffusive at these length scales, any difference in the measured thermal conductivity can only be explained by the presence of additional heat conduction channels.
Thus, by conducting the measurements for different heater-sensor distances, membrane thicknesses, and temperatures, we demonstrate how heat conduction by SPhPs occurs in thin membranes and remains quasi-ballistic for over hundreds of micrometres.

\section*{RESULTS}
\subsection*{Fabrication and measurements}

\begin{figure}[ht]
    \centering
    \includegraphics[width=1.0\linewidth]{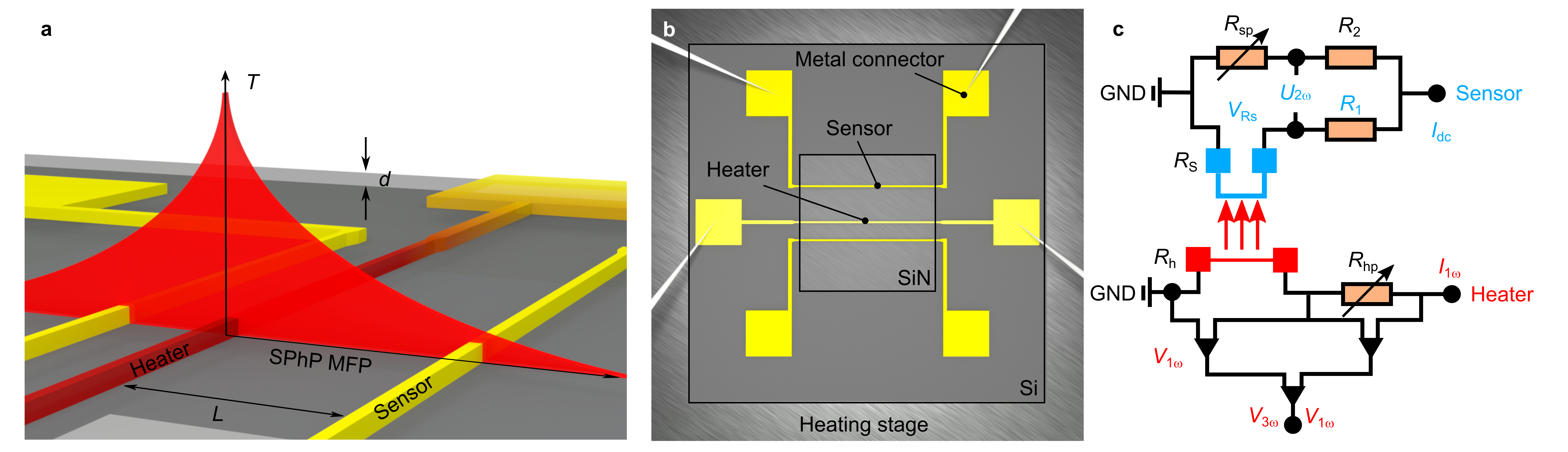}
    \caption{\textbf{Schemes of the $3\omega$ experimental setup.}
    (\textbf{a}) Illustration of a SiN membrane supporting the heater-to-sensor propagation of SPhPs excited by the Joule effect in the heater resistor. (\textbf{b}) Top view of a suspended SiN sample with a micro heater and two sensors. (\textbf{c}) Circuit diagram of the 3$\omega$ setup implemented with a heater $R_\text{H}$ and sensor $R_\text{S}$.}
    \label{Figure1}
\end{figure}

Samples of amorphous SiN membranes with thicknesses of 30, 50, 100, and 200 nm were suspended in a $1 \times 1$ mm$^2$ square windows on Si substrates as shown in Fig. \ref{Figure1}a. 
These high-stress ($\approx$ 250 MPa) membranes were flat (curvature radius of 4 m) to ensure an ideal measurement for the long-range SPhP propagation. 
Gold wires of 4 $\mu$m in width and 100 nm in height, serving the heater and the sensors (Supplementary Fig. S1), were deposited on the membranes using laser lithography and electron beam assisted physical deposition (Ulvac EX-300). 
A layer of 10-nm-thick Cr was deposited between Au and SiN to enhance the adhesion. 

To measure the in-plane thermal conductivity of SiN membranes, we used the 3$\omega$ method \cite{cahill1987thermal,cahill1990thermal,corbino1910oscillazioni}. 
Two sensors placed in parallel at 100 and 200 $\mu$m from the heater were deposited to quantify the ballistic heat transport of SPhPs for different propagation distances. 
Measurements with each sensor were carried out separately. 
Fig. \ref{Figure1}a shows the scheme of the used $3\omega$ setup with four probes coupled to a heating stage and placed in a vacuum chamber. 
SPhPs were excited by the Joule effect in the heater resistor. 
The in-plane thermal conductivity was extracted by comparing the 3$\omega$ signal with an analytical model (Supplementary Note 2). 
Thermal radiation losses in all SiN membranes were taken into account as explained in Supplementary Note 2.

As shown in Figs. \ref{Figure1}b and \ref{Figure1}c, the in-plane thermal conductivity measurements were based on 3$\omega$ method with a low-noise AC source (Lakeshore AC155) supplying the modulated current $I=I_0\text{cos}(\omega t)=I_0\text{Re}(\text{e}^{i\omega t})$ in a thin Cr/Au micro-resistor.
The supplied current caused the temperature increase d$T_{\text{h}}$ calculated from the $3\omega$ voltage measured with a lock-in amplifier in a half-bridge configuration.
A programmable analogue resistance $R_\text{hp}$ canceled the excessive 1$\omega$ signal at the input of the lock-in, enabling a higher sensitivity range for detecting the 3$\omega$ signal.
The heat flux propagated from the heater to the sensor resistance $R_\text{s}$, leading to an elevation of temperature d$T_\text{s}$ that was calculated from the $2\omega$ signal detected with another lock-in amplifier in a DC Wheatstone bridge. 
All analogue resistances had a Temperature Coefficient of Resistance (TCR) below 5 ppm/K. 
The holder temperature was controlled from 300 to 400 K with a Lakeshore 335. 
The measurements were operated in a vacuum probe station (JANIS ST-500-1) with a typical vacuum level of $5 \times 10^{-6}$ mbar (Pfeiffer Hi-Cube 80). 

A Labview program was developed to fully automate the parameter sweeps (holder temperature, heater frequency $w/2\pi$, heater amplitude $I_{1\omega}$, sensor DC $I_{\text{dc}}$) with an auto-adjustment of lock-in amplifiers sensitivity according to the signals levels. 
These time-dependent signals were continuously recorded for ten minutes for each experimental condition, except in the case of holder temperature changes, where the stabilization time was one hour.
Preliminary annealing at 450 K for 12 hours was performed in a vacuum before the measurements.
Such high thermal budget annealing allowed stabilizing the intrinsic properties of the Cr/Au micro-resistances and contribute to linear TCRs obtained for both $R_{\text{h}}$ and $R_{\text{s}}$, typically 1000 times higher than the TCR of any analogue resistance in the circuits.

\subsection*{SPhP propagation over hundreds of micrometres}

\begin{figure}[ht]
\centering
\includegraphics[width=0.5\linewidth]{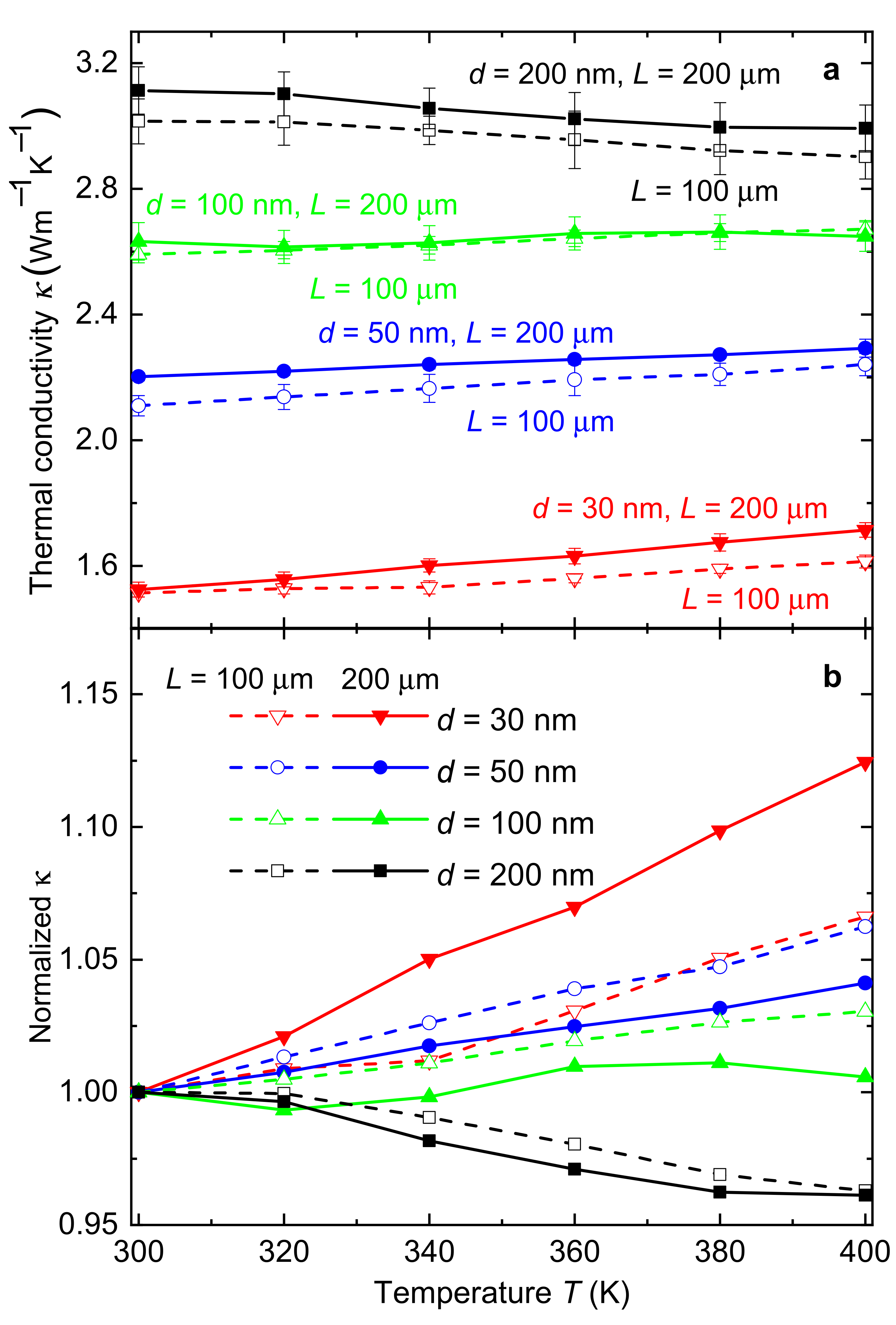}
\caption{\textbf{Thermal conductivity of SiN membranes.} (\textbf{a}) Thermal conductivity as a function of temperature for membranes of different thickness. The solid and dashed lines stand for the measurements performed with the sensors placed 100 and 200 $\mu$m away from the heater, respectively. (\textbf{b}) In-plane thermal conductivity normalized by its values at 300 K.}
\label{Figure3}
\end{figure}

Figure \ref{Figure3}a reports the measured temperature dependence of in-plane thermal conductivity of SiN membranes of different thickness.
The data measured for the heater-sensor distances of 100 and 200 $\mu$m are joined by the dashed and solid lines and labelled as $\kappa_{100}$ and $\kappa_{200}$, respectively. 
The thickest membrane ($d$ = 200 nm) at 300 K has $\kappa_{100} \approx \kappa_{200} \approx 3.1$ Wm$^{-1}$K$^{-1}$, which decreases by about 4\% at 400 K. 
This value of the thermal conductivity is consistent with our previous measurements \cite{wu2020enhanced}, and its reduction above room temperature is characteristic of the phonon thermal conductivity driven by the internal scattering processes of phonons.
Likewise, the thermal conductivity of the 100-nm-thick membrane is nearly independent of the heater-sensor distance and temperature.
The fact that the thermal conductivity of thicker membranes is independent of the heater-sensor distance indicates that the measured thermal conduction is mainly driven by the phonons and thus is diffusive at this scale, as predicted by theory \cite{ordonez-miranda_anomalous_2013}.

However, the thermal conductivity of the two thinner membranes (30 and 50 nm) increases with temperature.
For the 50-nm-thick membrane, $\kappa_{100}$ ($\kappa_{200}$) increases from 2.10 Wm$^{-1}$K$^{-1}$ (2.20 Wm$^{-1}$K$^{-1}$) at 300 K to 2.24 Wm$^{-1}$K$^{-1}$ (2.29 Wm$^{-1}$K$^{-1}$) at 400 K, which represents an enhancement of about 5\%, as shown in Fig. \ref{Figure3}b. 
For the 30-nm-thick membrane, this enhancement raises to 6.6\% and 12.4\% for the thermal conductivity values $\kappa_{100}$ and $\kappa_{200}$. 
The observed increase in the thermal conductivity with temperature is the sign of the heat conduction by SPhPs, as shown in our previous work \cite{wu2020enhanced}.
As SPhP contribution becomes stronger at higher temperatures, it compensates for the thermal conductivity reduction due to the phonon-phonon scattering, thus leading to the overall increase of the thermal conductivity with temperature.
This explains the difference in trends for thicker and thinner membranes.

Moreover, we observe different thermal conductivity for the different heater-sensor distances.
The thermal conductivity measured 200 $\mu$m away from the heater appears to be higher than that measured 100 $\mu$ away ($\kappa_{200} > \kappa_{100}$).
Since the phonon transport is diffusive at such a long length scale, we attribute the observed difference in the thermal conductivity to the ballistic heat conduction by SPhPs.
Indeed, length-dependent thermal conductivity is one of the signs of ballistic heat conduction \cite{anufriev2019probing, anufriev_quasi-ballistic_2018, zhang2018thermal, vakulov2020ballistic} and implies that heat carriers propagate ballistically at least as far as length dependence is measured.
Thus, the observations suggest that SPhPs can ballistically carry heat for at least 200 $\mu$m.

\section*{DISCUSSION}

To better understand the behaviour of the measured thermal conductivity, we analyze the SPhP contribution to the heat transport in SiN membranes. 
According to the Boltzmann transport equation, under the relaxation time approximation, the SPhP contribution ($\kappa_{\text{SPhP}}$) to the in-plane thermal conductivity of a membrane of thickness $d$ is given by\cite{chen_surface_2005}:

\begin{equation}
\label{TC}
 \kappa_{\text{SPhP}} = \frac{1}{4\pi d}\int_{\omega_\text{L}}^{\omega_\text{H}}\hslash\omega\Lambda_\text{e}\beta_\text{R}\frac{\partial f_0}{\partial T}\text{d}\omega,
\end{equation}

where $\hslash$ is the reduced Planck constant, $\Lambda_\text{e}$ is the effective propagation length of SPhPs propagating along the membrane surface with a complex wavevector $\beta$, $\beta_\text{R}=\text{Re}(\beta)$, $f_0$ is the Bose-Einstein distribution function, $T$ is the average membrane temperature, and $\omega_\text{H}$ and $\omega_\text{L}$ stand for the highest and lowest frequencies supporting the propagation of SPhPs, respectively. 
Taking into account that the heat generated by SPhPs inside material is proportional to the imaginary part of its dielectric function \cite{baffou_mapping_2010}, this latter material property strongly determines these frequencies and hence $\kappa_{\text{SPhP}}$. 

\begin{figure}[ht]
\centering
\includegraphics[width=1\linewidth]{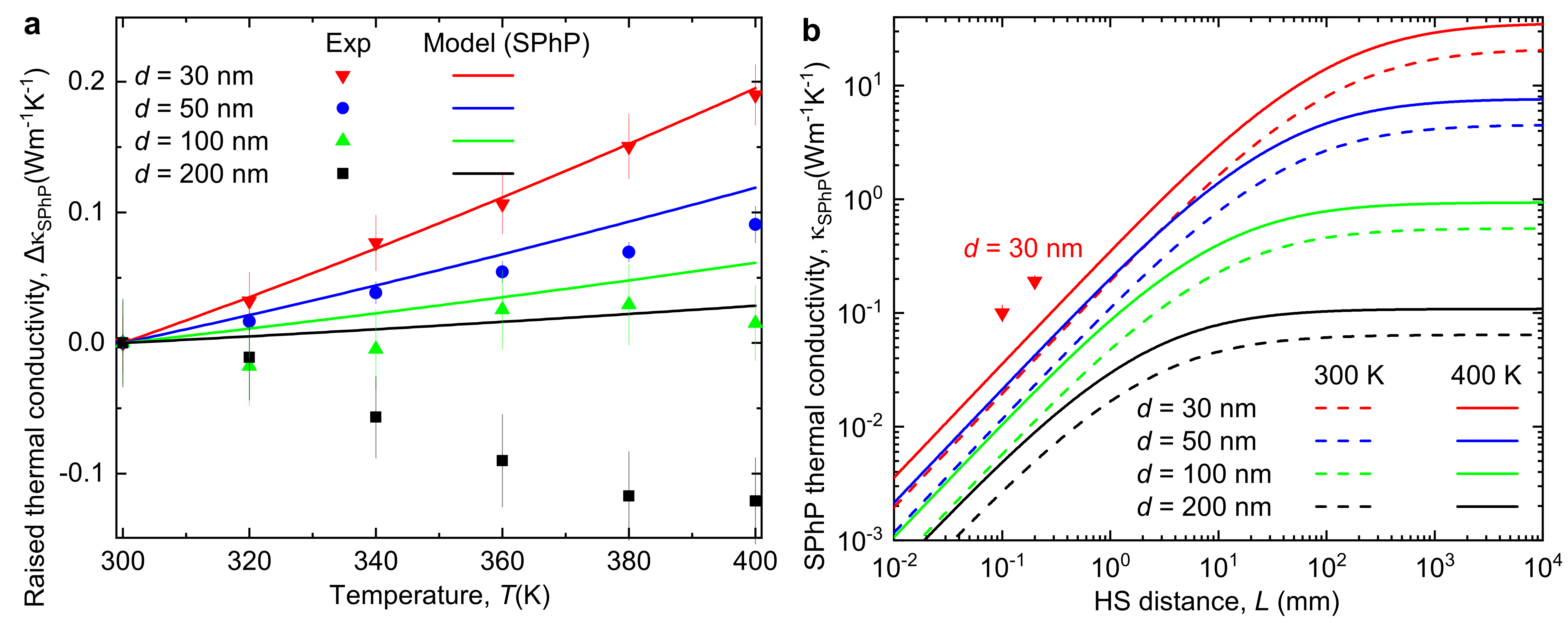}
\caption{\textbf{Theoretical SPhP thermal conductivity agrees with the experimental ones.} (\textbf{a}) Enhanced SPhP thermal conductivity as a function of temperature. (\textbf{b}) Theoretical SPhP thermal conductivity as a function of the heater-sensor distance for four SiN membrane thicknesses and for 300 and 400 K. The experimental result of 30-nm-thick membrane (red solid triangles) is in good agreement with theoretical values.}
\label{Figure4}
\end{figure}

To take into account the physical limitations of our SiN membranes for the propagation of SPhPs, their effective thermal conductivity is determined by Mathiessen’s rule $\Lambda_\text{e}^{-1}=\Lambda^{-1}+L^{-1}$, where $\Lambda$ is the intrinsic propagation length of polaritons, and $L$ is the heater-sensor distance (100 and 200 $\mu$m). 
Thus, the values of $L$ are much shorter than those of $\Lambda$ ($\approx$ 1 m for SiN membranes \cite{wu2020enhanced}) and much longer than the intrinsic mean free path of phonons in amorphous materials \cite{pan2020quantifying, wingert_thermal_2016} around room temperature. 
Equation \ref{TC} thus establishes that the values of $\kappa_{\text{SPhP}}$ are determined by the SPhP dispersion relation $\beta(\omega)$ driven by the membrane dielectric function that usually changes with temperature and membrane thickness\cite{joulain_radiative_2015}. 
However, for the thicknesses and temperatures considered in this work, the dielectric function of our SiN membranes is nearly independent of these parameters, as established by our previous optical measurements \cite{wu2020enhanced}.

The SPhP thermal conductivity predicted by Eq. \ref{TC} as a function of temperature and heater-sensor distance is respectively shown in Figs. \ref{Figure4}a and \ref{Figure4}b for different SiN membrane thicknesses.
Thinner and hotter membranes with longer heater-sensor distances exhibit higher SPhP thermal conductivity.
This behaviour is consistent with that reported for SiO$_2$ and SiC membranes \cite{chen_surface_2005, ordonez-miranda_anomalous_2013} and opposite to the typical behaviour of the phonon thermal conductivity.
Thus, it signals the SPhP contribution to the heat transport along the membranes.

Fig. \ref{Figure4}a shows that for the thinnest membrane ($d$ = 30 nm) the measured thermal conductivity enhancement $\Delta\kappa_{\text{SPhP}} = \kappa(T) - \kappa(300 K)$ is in agreement with prediction of the model. 
However, in thicker membranes ($d >$ 30 nm), the phonons make a significant contribution to heat conduction, and the discrepancy between measured (phonons and SPhPs) and theoretical (only SPhPs) values increases with membrane thickness. 
Remarkably, Fig. \ref{Figure4}b shows that theoretically, SPhPs can travel ballistically as far as one meter, and the values measured on the 30-nm-thick membrane follow the predicted trend.

In conclusion, we experimentally demonstrated long-distance heat conduction by SPhPs.
We measured the in-plane thermal conductivity of SiN membranes and showed that in thin membranes, the heat conduction is mainly driven by SPhPs.
Moreover, the measured thermal conductivity was higher when measured further away from the heater.
The observed difference in the thermal conductivity is attributed to ballistic heat conduction by SPhPs and is in agreement with our theoretical model.
Thus, our experimental results suggest that the SPhPs can carry heat quasi-ballistically over hundreds of micrometres, whereas our model predicts that it is far from the limit. 
This work lays the foundations for improving the heat dissipation in microelectronics and efficiency in silicon photonics.

\section*{Acknowledgement}
\subsection*{Funding}
This work is supported by CREST JST Grant Number JPMJCR19Q3 and JPMJCR19I1, KAKENHI Grant Number 21H04635.

\subsection*{Author contributions}
YW, LJ designed the experiment. 
YW fabricated the samples and wrote the paper.
RA and ST contributed to the fabrication of the samples.
LJ and HF initialized the 3$\omega$ setup.
LJ performed the 3$\omega$ measurements.
JOM developed the theoretical and analytical models.
JOM and RA contributed to writing the paper.
SV and MN contributed to the interpretation of the results and supervised the entirety of the work.
All authors contributed to the analysis, discussion, and preparation of the manuscript.

\subsection*{Competing interests}
The authors declare that they have no competing interests.
\subsection*{Data and materials availability}
All data needed to evaluate the conclusions in the paper are present in the article and the Supplementary Materials. 
Additional data related to this paper may be requested from the authors. Correspondence and requests for materials should be addressed to YW (yunhui@iis.u-tokyo.ac.jp) or MN (nomura@iis.u-tokyo.ac.jp).

\bibliography{reference.bib}
\end{document}